\documentclass[10pt]{article}
\usepackage{graphicx}
\usepackage{amsmath}
\usepackage{amssymb}
\usepackage{caption2}
\begin{document}
\bibliographystyle{prsty}
\begin{center}
{\large {\bf \sc{  Revisit the tetraquark candidates in the   $J/\psi J/\psi$ mass spectrum }}} \\[2mm]
Zhi-Gang  Wang \footnote{E-mail: zgwang@aliyun.com.  }   \\
 Department of Physics, North China Electric Power University, Baoding 071003, P. R. China
\end{center}

\begin{abstract}
In this article, we introduce a relative P-wave to construct the doubly-charm axialvector diquark operator, then take the doubly-charm axialvector (anti)diquark operator as the basic constituent to construct the scalar and tensor tetraquark currents to study the scalar, axialvector and tensor fully-charm tetraquark states with the QCD sum rules. We observe that the  ground state  $\tilde{A}\tilde{A}$ type tetraquark states  and the first radial excited states of the $AA$ type tetraquark states have almost degenerated masses, where the $\tilde{A}$ and $A$ stand for the diquark operators with and without the relative P-wave respectively,  the broad structure above the $J/\psi J/\psi$ threshold  maybe consist of several  diquark-antidiquark type fully-charm  tetraquark states.
\end{abstract}



\section{Introduction}
 In the constituent quark models, we usually classify the hadrons into conventional mesons and baryons, and exotic tetraquark states,  pentaquark states and hexaquark states, etc. The $X(3872)$, the first exotic candidate  observed in 2003 by the Belle collaboration \cite{X3872-2003},  has hidden-charm, but cannot be
fitted into any radial or orbital excitation of the charmonium, it should have more complicated structure than a mere $c\bar{c}$ pair.
 The exotic states provide a unique environment to explore  the strong interaction, which governs the dynamics of the quarks and gluons, and the confinement mechanism.
All the hadrons listed in {\it The Review of Particle Physics } to date contain  two heavy  valence quarks at most \cite{PDG}, whereas many QCD-motivated phenomenological models  permit
the existence of tetraquark states consisting of four heavy valence quarks,  the fully-heavy tetraquark states have attracted much attentions in recent years and have been studied extensively  \cite{Lloyd-2004,WZG-QQQQ-EPJC,Rosner-2017,ChenW-PLB-2017,Hughes-2018,Navarra-2019}.

Recently, the LHCb  collaboration studied the $J/\psi J/\psi$   invariant mass spectrum using proton-proton collision data at centre-of-mass energies of $\sqrt{s}=$7, 8  and 13 TeV  recorded by the LHCb experiment  corresponding to an integrated luminosity of $9\,\rm{fb}^{-1}$, and observed a narrow resonance structure around $6.9\, \rm{GeV}$  and   a broad structure just above the $J/\psi J/\psi$ mass with global significances of more than five standard deviations \cite{LHCb-cccc-2006}. The Breit-Wigner mass and width of the $X(6900)$ are
\begin{eqnarray}
M_X&=&6905 \pm 11 \pm 7\, \rm{MeV} \, , \nonumber \\
\Gamma_{X}&=&80 \pm 19 \pm  33\, \rm{MeV} \, ,
\end{eqnarray}
assuming no interference with the nonresonant single-parton scattering continuum. When assuming the nonresonant single-parton scattering continuum interferes
with the broad structure close to the $J/\psi J/\psi$  mass threshold, the Breit-Wigner mass and width are changed to
\begin{eqnarray}
M_X&=&6886 \pm  11 \pm 11\rm{MeV} \, , \nonumber \\
\Gamma_{X}&=&168 \pm  33 \pm  69\, \rm{MeV} \, .
\end{eqnarray}

Both the narrow and broad resonance structures are observed in the $J/\psi J/\psi$ invariant mass spectrum, such structures  are
naturally assigned to have  the valence quarks or constituent quarks  $c\bar c c\bar c$,
which makes them the first fully-heavy exotic multiquark candidates claimed  experimentally to date, their observations  have revitalized the investigations  of  multiquark resonances made of heavy quarks and heavy antiquarks \cite{Zhong-2020-P-tetraquark,WZG-QQQQ-CPC,PingJL-QQQQ,ZhangJR-QQQQ}.
It is  a very important step in investigations   of the heavy hadrons, after discoveries  of the charmonium $c\bar c$ in 1974, and the charmed mesons $c\bar q$ and baryons $cqq$ in the subsequent years;  the bottomonium $b\bar b$ in 1977, and the bottom mesons $b\bar q$ and baryons $bqq$ in the subsequent years; the $B_c$ in 1996 at  the Fermilab Tevatron collider; and the double-charm baryons $ccq$ in 2017 by the LHCb collaboration \cite{PDG}.

In spite of a large body of experimental information   accumulated on the  exotic hadrons, we have never reached  consensus  on the way  the valence quarks are organized inside them, the diquark-antidiquark type, color-singlet-color-singlet type, or other type quark structures?
  In the present case, there are no known color-singlet light mesons can  be exchanged between two charmonium states to produce binding energies or
  final-state interactions.
 The thresholds of the charmonium pairs $\eta_c \eta_c$, $J/\psi J/\psi$, $\chi_{c0}\chi_{c0}$, $\chi_{c1}\chi_{c1}$, $h_ch_c$ and $\chi_{c2}\chi_{c2}$ are $5.97\,\rm{GeV}$, $6.19\,\rm{GeV}$, $6.83\,\rm{GeV}$, $7.02\,\rm{GeV}$, $7.05\,\rm{GeV}$ and $7.11\,\rm{GeV}$ respectively from the Particle Data Group \cite{PDG}.
 The $X(6900)$ lies about $700\,\rm{MeV}$ above the $J/\psi J/\psi$ threshold, and  in the
vicinity of the $\chi_{c0} \chi_{c0}$ and $\chi_{c1} \chi_{c0}$
thresholds, it is very difficult   to produce such a strong resonance structure through the threshold rescattering mechanism.
 As a result, the most general  models for the fully-heavy four-quark  states resort to the diquark-antidiquark
 configurations,  the attractive interactions between the two heavy quarks or antiquarks should dominate at the short distance and favor
forming the genuine diquark-antidiquark type tetraquark states rather
than the loosely-bound tetraquark molecular states.
 Needless to say, determining the spin-parity of the resonances is in the first priority.

More precisely, the attractive (repulsive) interaction in the antisymmetric (symmetric) color  antitriplet (sextet) channel originates from the one-gluon exchange   favors (disfavors)  formation of the diquarks in the color antitriplet (sextet). We prefer the  diquarks in the color antitriplet to the diquarks in the color sextet in constructing the tetraquark current operators.
The  diquark operators $\varepsilon^{ijk} q^{T}_j C\Gamma q^{\prime}_k$  have  five  structures  in Dirac spinor space, where the $i$, $j$ and $k$ are color indexes, $C\Gamma=C\gamma_5$, $C$, $C\gamma_\mu \gamma_5$,  $C\gamma_\mu $ and $C\sigma_{\mu\nu}$ for the scalar, pseudoscalar, vector, axialvector  and  tensor diquarks, respectively.
The  favorite   diquark configurations are the scalar ($C\gamma_5$) and axialvector ($C\gamma_\mu$) diquark states from the QCD sum rules \cite{WangDiquark,WangLDiquark}.
 The double-heavy diquark operators  $\varepsilon^{ijk} Q^{T}_j C\gamma_5 Q_k$ cannot exist due to the Fermi-Dirac statistics.
 In previous work, we took the doubly-heavy diquark operators $\varepsilon^{ijk} Q^{T}_j C\gamma_\mu Q_k$ ($A$) as basic constituents to construct the scalar and tensor currents to study the scalar, axialvector, vector, tensor tetraquark states  and their radial excited states   with the QCD sum rules \cite{WZG-QQQQ-EPJC,WZG-QQQQ-CPC}.
Now we introduce the explicit P-wave to construct the axialvector doubly-heavy diquark operators $\varepsilon^{ijk} Q^{T}_j C\gamma_5\stackrel{\leftrightarrow}{\partial}_\mu Q_k$ ($\tilde{A}$), which can exist due  to the Fermi-Dirac statistics,  the derivative  $\stackrel{\leftrightarrow}{\partial}_\mu=\stackrel{\rightarrow}{\partial}_\mu-\stackrel{\leftarrow}{\partial}_\mu$ embodies  the P-wave effect.

In this article, we take the axialvector diquark operator $\tilde{A}$ as the basic constituent, construct  the $\tilde{A}\tilde{A}$ type scalar and tensor tetraquark currents to study the mass spectrum of the ground states of the scalar, axialvector and tensor fully-charm  tetraquark states with the QCD sum rules,  and try to make possible assignments of the LHCb's new resonance structures.

The article is arranged as follows:  we derive the QCD sum rules for the masses and pole residues of the  $cc\bar{c}\bar{c}$ tetraquark states in section 2; in section 3, we present the numerical results and discussions; section 4 is reserved for our conclusion.

\section{QCD sum rules for  the  $\tilde{A}\tilde{A}$ type  tetraquark states  }
We  write down  the two-point correlation functions   $\Pi (p)$ and $\Pi_{\mu\nu\alpha\beta}(p)$ in the QCD sum rules firstly,
\begin{eqnarray}
\Pi(p)&=&i\int d^4x e^{ip \cdot x} \langle0|T\left\{J(x)J^{\dagger}(0)\right\}|0\rangle \, ,\nonumber\\
\Pi_{\mu\nu\alpha\beta}(p)&=&i\int d^4x e^{ip \cdot x} \langle0|T\left\{J_{\mu\nu}(x)J_{\alpha\beta}^{\dagger}(0)\right\}|0\rangle \, ,
\end{eqnarray}
where $J_{\mu\nu}(x)=J^1_{\mu\nu}(x)$, $J^2_{\mu\nu}(x)$,
\begin{eqnarray}
J(x)&=&\varepsilon^{ijk}\varepsilon^{imn}c^{Tj}(x)C\gamma_5\stackrel{\leftrightarrow}{\partial}_\mu c^k(x)\, \bar{c}^m(x)\stackrel{\leftrightarrow}{\partial}_\nu \gamma_5C \bar{c}^{Tn}(x) \, g^{\mu\nu}\, , \nonumber\\
J^1_{\mu\nu}(x)&=&\varepsilon^{ijk}\varepsilon^{imn}\Big\{c^{Tj}(x)C\gamma_5\stackrel{\leftrightarrow}{\partial}_\mu c^k(x)\, \bar{c}^m(x) \stackrel{\leftrightarrow}{\partial}_\nu \gamma_5C \bar{c}^{Tn}(x)\nonumber\\
&&-c^{Tj}(x)C\gamma_5\stackrel{\leftrightarrow}{\partial}_\nu  c^k(x)\, \bar{c}^m(x)\stackrel{\leftrightarrow}{\partial}_\mu \gamma_5C \bar{c}^{Tn}(x) \Big\} \, , \nonumber \\
J^2_{\mu\nu}(x)&=&\varepsilon^{ijk}\varepsilon^{imn}\Big\{c^{Tj}(x)C\gamma_5\stackrel{\leftrightarrow}{\partial}_\mu c^k(x) \,\bar{c}^m(x) \stackrel{\leftrightarrow}{\partial}_\nu \gamma_5C \bar{c}^{Tn}(x)\nonumber\\
&&+c^{Tj}(x)C\gamma_5\stackrel{\leftrightarrow}{\partial}_\nu  c^k(x)\, \bar{c}^m(x)\stackrel{\leftrightarrow}{\partial}_\mu \gamma_5C \bar{c}^{Tn}(x) \Big\} \, ,
\end{eqnarray}
 the $i$, $j$, $k$, $m$, $n$ are color indexes, the $C$ is the charge conjugation matrix. We choose  the tetraquark   currents $J(x)$, $J^1_{\mu\nu}(x)$ and $J^2_{\mu\nu}(x)$ to interpolate the $J^{PC}=0^{++}$, $1^{+-}$ and $2^{++}$ diquark-antidiquark type  tetraquark states, respectively.

At the hadron  side, we  insert  a complete set of intermediate hadronic states with
the same quantum numbers as the tetraquark current operators $J(x)$, $J^1_{\mu\nu}(x)$ and $J^2_{\mu\nu}(x)$ into the
correlation functions $\Pi(p)$ and $\Pi_{\mu\nu\alpha\beta}(p)$ to obtain the hadronic representation
\cite{SVZ79,Reinders85}. After isolating the ground state
contributions of the scalar, axialvector  and tensor fully-charm   tetraquark states, we obtain the results,
\begin{eqnarray}
\Pi (p) &=&\frac{\lambda_X^2}{M^2_X-p^2} +\cdots \, \, , \nonumber\\
&=&\Pi_S(p^2)\, ,
\end{eqnarray}
\begin{eqnarray}
\Pi^1_{\mu\nu\alpha\beta}(p)&=&\frac{\lambda_{ Y^+}^2}{M_{Y^+}^2\left(M_{Y^+}^2-p^2\right)}\left(p^2g_{\mu\alpha}g_{\nu\beta} -p^2g_{\mu\beta}g_{\nu\alpha} -g_{\mu\alpha}p_{\nu}p_{\beta}-g_{\nu\beta}p_{\mu}p_{\alpha}+g_{\mu\beta}p_{\nu}p_{\alpha}+g_{\nu\alpha}p_{\mu}p_{\beta}\right) \nonumber\\
&&+\frac{\lambda_{ Y^-}^2}{M_{Y^-}^2\left(M_{Y^-}^2-p^2\right)}\left( -g_{\mu\alpha}p_{\nu}p_{\beta}-g_{\nu\beta}p_{\mu}p_{\alpha}+g_{\mu\beta}p_{\nu}p_{\alpha}+g_{\nu\alpha}p_{\mu}p_{\beta}\right) +\cdots \, \, ,\nonumber\\
&=&\widetilde{\Pi}_{A}(p^2)\left(p^2g_{\mu\alpha}g_{\nu\beta} -p^2g_{\mu\beta}g_{\nu\alpha} -g_{\mu\alpha}p_{\nu}p_{\beta}-g_{\nu\beta}p_{\mu}p_{\alpha}+g_{\mu\beta}p_{\nu}p_{\alpha}+g_{\nu\alpha}p_{\mu}p_{\beta}\right) \nonumber\\
&&+\widetilde{\Pi}_{V}(p^2)\left( -g_{\mu\alpha}p_{\nu}p_{\beta}-g_{\nu\beta}p_{\mu}p_{\alpha}+g_{\mu\beta}p_{\nu}p_{\alpha}+g_{\nu\alpha}p_{\mu}p_{\beta}\right) \, .
\end{eqnarray}
\begin{eqnarray}
\Pi^2_{\mu\nu\alpha\beta} (p) &=&\frac{\lambda_X^2}{M_X^2-p^2}\left( \frac{\widetilde{g}_{\mu\alpha}\widetilde{g}_{\nu\beta}+\widetilde{g}_{\mu\beta}\widetilde{g}_{\nu\alpha}}{2}-\frac{\widetilde{g}_{\mu\nu}\widetilde{g}_{\alpha\beta}}{3}\right) +\cdots \, \, ,  \nonumber\\
&=&\Pi_{T}(p^2)\left( \frac{\widetilde{g}_{\mu\alpha}\widetilde{g}_{\nu\beta}+\widetilde{g}_{\mu\beta}\widetilde{g}_{\nu\alpha}}{2}-\frac{\widetilde{g}_{\mu\nu}\widetilde{g}_{\alpha\beta}}{3}\right) +\cdots \, \, ,
\end{eqnarray}
where $\widetilde{g}_{\mu\nu}=g_{\mu\nu}-\frac{p_{\mu}p_{\nu}}{p^2}$, the pole residues  $\lambda_{X}$ and $\lambda_{Y}$ are defined by
\begin{eqnarray}
 \langle 0|J (0)|X (p)\rangle &=& \lambda_{X}     \, , \nonumber\\
  \langle 0|J^1_{\mu\nu}(0)|Y^+(p)\rangle &=& \frac{\lambda_{Y^+}}{M_{Y^+}} \, \varepsilon_{\mu\nu\alpha\beta} \, \varepsilon^{\alpha}p^{\beta}\, , \nonumber\\
 \langle 0|J^1_{\mu\nu}(0)|Y^-(p)\rangle &=& \frac{\lambda_{Y^-}}{M_{Y^-}} \left(\varepsilon_{\mu}p_{\nu}-\varepsilon_{\nu}p_{\mu} \right)\, ,\nonumber\\
  \langle 0|J^2_{\mu\nu}(0)|X (p)\rangle &=& \lambda_{X} \, \varepsilon_{\mu\nu}   \, ,
\end{eqnarray}
the superscripts $\pm$ on the $Y$ stand for the parity of the tetraquark states, the $\varepsilon_{\mu}$ and $\varepsilon_{\mu\nu} $ are the  polarization vectors of the axialvector, vector  and tensor tetraquark states, respectively.
In Ref.\cite{WangZG-IJMPA-Z3900},  we assign the $Z_c(3900)$ to be an axialvector tetraquark state tentatively, and study it  with the QCD sum rules in details by including the two-particle scattering state contributions   and nonlocal effects between the diquark and antidiquark constituents. In calculations, we observe that  the two-particle scattering state contributions cannot saturate  the QCD sum rules at the hadron side,  the contribution of the $Z_c(3900)$ plays an un-substitutable role, we can saturate the QCD sum rules with or without the two-particle scattering state contributions. The conclusion is applicable in the present case, and we neglect the contributions of the intermediate charmonium pairs, such as $\eta_c \eta_c$, $J/\psi J/\psi$, $\chi_{c0}\chi_{c0}$, etc.

We project out the axialvector and vector components $\widetilde{\Pi}_{A}(p^2)$ and $\widetilde{\Pi}_{V}(p^2)$ by introducing the operators $P_{A}^{\mu\nu\alpha\beta}$ and $P_{V}^{\mu\nu\alpha\beta}$, respectively,
\begin{eqnarray}
\Pi_{A}(p^2)&=&p^2\widetilde{\Pi}_{A}(p^2)=P_{A}^{\mu\nu\alpha\beta}\Pi_{\mu\nu\alpha\beta}(p) \, , \nonumber\\
\Pi_{V}(p^2)&=&p^2\widetilde{\Pi}_{V}(p^2)=P_{V}^{\mu\nu\alpha\beta}\Pi_{\mu\nu\alpha\beta}(p) \, ,
\end{eqnarray}
where
\begin{eqnarray}
P_{A}^{\mu\nu\alpha\beta}&=&\frac{1}{6}\left( g^{\mu\alpha}-\frac{p^\mu p^\alpha}{p^2}\right)\left( g^{\nu\beta}-\frac{p^\nu p^\beta}{p^2}\right)\, , \nonumber\\
P_{V}^{\mu\nu\alpha\beta}&=&\frac{1}{6}\left( g^{\mu\alpha}-\frac{p^\mu p^\alpha}{p^2}\right)\left( g^{\nu\beta}-\frac{p^\nu p^\beta}{p^2}\right)-\frac{1}{6}g^{\mu\alpha}g^{\nu\beta}\, .
\end{eqnarray}
The vector tetraquark state $Y^-$ has negative parity, and should have an additional P-wave compared to the tetraquark states $X$ with the positive parity, and is beyond the present work as there are three P-waves.

It is straightforward but tedious to compute  the operator product expansion in  the deep Euclidean  space $P^2=-p^2 \to \infty$, then we obtain the QCD spectral densities through dispersion relation,
\begin{eqnarray}
\Pi_{S/A/T}(p^2)&=& \int_{16m_c^2}^{\infty}ds \frac{\rho_{S/A/T}(s)}{s-p^2}\, ,
\end{eqnarray}
where
\begin{eqnarray}
\rho_{S/A/T}(s)&=&\frac{{\rm Im}\Pi_{S/A/T}(s)}{\pi}\, .
\end{eqnarray}

 We  take the quark-hadron duality below the continuum thresholds  $s_0$,  and perform Borel transform  in regard to
the variable $P^2=-p^2$ to obtain  the QCD sum rules:
\begin{eqnarray}\label{QCDSR}
\lambda^2_{X/Y}\, \exp\left(-\frac{M^2_{X/Y}}{T^2}\right)&=& \int_{16m_c^2}^{s_0} ds \int_{4m_c^2}^{\left(\sqrt{s}-2m_c\right)^2}dt \int_{4m_c^2}^{\left(\sqrt{s}-\sqrt{t}\right)^2}dr \, \rho(s,t,r)  \exp\left(-\frac{s}{T^2}\right) \, ,
\end{eqnarray}
where the QCD spectral densities  $\rho(s,t,r) =\rho_S(s,t,r) $, $\rho_A(s,t,r)$ and $\rho_T(s,t,r) $,

\begin{eqnarray}
 \rho_S(s,t,r)&=&\frac{\sqrt{\lambda(s,t,r)\,\lambda(t,m_c^2,m_c^2)\,\lambda(r,m_c^2,m_c^2)}}{3072\pi^6}\left(1-\frac{4m_c^2}{t} \right)\left(1-\frac{4m_c^2}{r} \right)\nonumber\\
 &&\left(s-2t-2r +\frac{r^2+t^2+10rt}{s} \right)\nonumber
 \end{eqnarray}
 \begin{eqnarray}
 &&+\langle\frac{\alpha_sGG}{\pi}\rangle\frac{\sqrt{\lambda(s,t,r)\,\lambda(r,m_c^2,m_c^2)}}{288\pi^4}\left(1-\frac{4m_c^2}{r}\right) \frac{m_c^2}{t-4m_c^2}\frac{1}{\sqrt{t\left(t-4m_c^2\right)}}\nonumber\\
 &&\left(4t+4r-2s-24m_c^2+\frac{12m_c^2(r+t)-2r^2-2t^2-2rt-15m_c^4}{s} \right. \nonumber\\
 &&\left.+\frac{12sm_c^2-24rm_c^2+30m_c^4}{t}+\frac{12r^2m_c^2-6rm_c^4}{st}+\frac{30rm_c^4-15sm_c^4}{t^2} -\frac{15r^2m_c^4}{st^2} \right)\nonumber\\
 &&+\langle\frac{\alpha_sGG}{\pi}\rangle\frac{\sqrt{\lambda(s,t,r)\,\lambda(r,m_c^2,m_c^2)}}{384\pi^4}\left(1-\frac{4m_c^2}{r}\right) \frac{m_c^2}{\sqrt{t\left(t-4m_c^2\right)}}\nonumber\\
 &&\left( -6+\frac{3t+6r-10m_c^2}{s}+\frac{3s-6r+20m_c^2}{t}+\frac{3r^2-4rm_c^2}{st}+\frac{20rm_c^2-10sm_c^2}{t^2}-\frac{10r^2m_c^2}{st^2}\right) \nonumber\\
 &&+\langle\frac{\alpha_sGG}{\pi}\rangle\frac{\sqrt{\lambda(s,t,r)\,\lambda(r,m_c^2,m_c^2)}}{256\pi^4}\left(1-\frac{4m_c^2}{r}\right) \frac{1}{\sqrt{t\left(t-4m_c^2\right)}}\nonumber\\
 &&\left( s-2t-2r+4m_c^2+\frac{r^2+t^2-14rt-2tm_c^2+28rm_c^2-20m_c^4}{s} +\frac{4rm_c^2-2sm_c^2+40m_c^4}{t}\right. \nonumber\\
 &&\left. -\frac{2r^2m_c^2+8rm_c^4}{st}+\frac{40rm_c^4-20sm_c^4}{t^2}-\frac{20r^2m_c^4}{st^2}\right)\, ,
\end{eqnarray}

\begin{eqnarray}
 \rho_{A}(s,t,r)&=&\frac{\sqrt{\lambda(s,t,r)\,\lambda(t,m_c^2,m_c^2)\,\lambda(r,m_c^2,m_c^2)}}{4608\pi^6}\left(1-\frac{4m_c^2}{t} \right)\left(1-\frac{4m_c^2}{r} \right)\nonumber\\
 &&\left(t+r-\frac{2r^2+2t^2-8rt}{s}+\frac{(t+r)(r-t)^2}{s^2} \right)\nonumber\\
 &&+\langle\frac{\alpha_sGG}{\pi}\rangle\frac{\sqrt{\lambda(s,t,r)\,\lambda(r,m_c^2,m_c^2)}}{864\pi^4}\left(1-\frac{4m_c^2}{r}\right) \frac{m_c^2}{t-4m_c^2}\frac{1}{\sqrt{t\left(t-4m_c^2\right)}}\nonumber\\
 &&\left(6m_c^2-4r-t+\frac{8r^2+2t^2-2rt-12tm_c^2+12rm_c^2+12m_c^4}{s}+\frac{24rm_c^2-6m_c^4}{t} \right. \nonumber\\
 &&\left. -\frac{48r^2m_c^2}{st} +\frac{7r^2t-t^3-4r^3-2rt^2+6t^2m_c^2-42r^2m_c^2+12rtm_c^2-6tm_c^4-18rm_c^4}{s^2}\right.\nonumber\\
 &&\left.-\frac{30rm_c^4}{t^2} +\frac{24r^3m_c^2+54r^2m_c^4}{s^2t}+\frac{60r^2m_c^4}{st^2}-\frac{30r^3m_c^4}{s^2t^2}\right) \nonumber\\
  &&+\langle\frac{\alpha_sGG}{\pi}\rangle\frac{\sqrt{\lambda(s,t,r)\,\lambda(r,m_c^2,m_c^2)}}{576\pi^4}\left(1-\frac{4m_c^2}{r}\right) \frac{m_c^2}{\sqrt{t\left(t-4m_c^2\right)}}\nonumber\\
  &&\left(1+\frac{4r-2t+4m_c^2}{s} +\frac{3r-2m_c^2}{t}+\frac{t^2-5r^2+rt-6rm_c^2-2tm_c^2}{s^2}-\frac{10rm_c^2}{t^2}\right.\nonumber\\
  &&\left. -\frac{6r^2}{st}+\frac{3r^3+18r^2m_c^2}{s^2t}+\frac{20r^2m_c^2}{st^2}-\frac{10r^3m_c^2}{s^2t^2}\right)\nonumber\\
 &&+\langle\frac{\alpha_sGG}{\pi}\rangle\frac{\sqrt{\lambda(s,t,r)\,\lambda(r,m_c^2,m_c^2)}}{384\pi^4}\left(1-\frac{4m_c^2}{r}\right) \frac{1}{\sqrt{t\left(t-4m_c^2\right)}}\nonumber\\
  &&\left(r-t+2m_c^2+\frac{2t^2-2r^2-12rt-4tm_c^2+24rm_c^2+8m_c^4}{s}-\frac{2rm_c^2+4m_c^4}{t}+\frac{4r^2m_c^2}{st}  \right. \nonumber\\
 &&\left. -\frac{20rm_c^4}{t^2} +\frac{40r^2m_c^4}{st^2}+\frac{(r-t)^3+2t^2m_c^2+6r^2m_c^2-6rtm_c^2-4tm_c^4-12rm_c^4}{s^2}\right. \nonumber\\
 &&\left.+\frac{36r^2m_c^4-2r^3m_c^2}{s^2t}-\frac{20r^3m_c^4}{s^2t^2} \right)\, ,
 \end{eqnarray}

\begin{eqnarray}
 \rho_{T}(s,t,r)&=&\frac{\sqrt{\lambda(s,t,r)\,\lambda(t,m_c^2,m_c^2)\,\lambda(r,m_c^2,m_c^2)}}{23040\pi^6}\left(1-\frac{4m_c^2}{t} \right)\left(1-\frac{4m_c^2}{r} \right)\nonumber\\
 &&\left(s+6t+6r-\frac{14r^2+14t^2-84rt}{s}+\frac{6(t+r)(r-t)^2}{s^2}+\frac{(r-t)^4}{s^3} \right)\nonumber\\
&&+\langle\frac{\alpha_sGG}{\pi}\rangle\frac{\sqrt{\lambda(s,t,r)\,\lambda(r,m_c^2,m_c^2)}}{2160\pi^4}\left(1-\frac{4m_c^2}{r}\right) \frac{m_c^2}{t-4m_c^2}\frac{1}{\sqrt{t\left(t-4m_c^2\right)}}\nonumber\\
&&\left(3t-12r-2s-18m_c^2+\frac{28r^2-2t^2-18rt+12tm_c^2+108rm_c^2-30m_c^4}{s} \right. \nonumber\\
&&\left.+\frac{12sm_c^2+72rm_c^2+30m_c^4}{t}-\frac{168r^2m_c^2+40rm_c^4}{st}-\frac{90rm_c^4+15sm_c^4}{t^2} \right. \nonumber\\
&&\left. +\frac{3t^3-12r^3+27r^2t-18rt^2-18t^2m_c^2-162r^2m_c^2+108rtm_c^2+30tm_c^4-150rm_c^4}{s^2}\right.\nonumber\\
&&\left.+\frac{12t^3m_c^2-2(t-r)^4-48r^3m_c^2+72r^2tm_c^2-48rt^2m_c^2+60rtm_c^4-15t^2m_c^4-90r^2m_c^4}{s^3} \right.\nonumber\\
&&\left.+\frac{72r^3m_c^2+210r^2m_c^4}{s^2t} +\frac{210r^2m_c^4}{st^2}+\frac{12r^4m_c^2+60r^3m_c^4}{s^3t}-\frac{90r^3m_c^4}{s^2t^2} -\frac{15r^4m_c^4}{s^3t^2}\right)\nonumber\\
&&+\langle\frac{\alpha_sGG}{\pi}\rangle\frac{\sqrt{\lambda(s,t,r)\,\lambda(r,m_c^2,m_c^2)}}{2880\pi^4}\left(1-\frac{4m_c^2}{r}\right) \frac{m_c^2}{\sqrt{t\left(t-4m_c^2\right)}}\nonumber\\
&&\left( -2+\frac{52r-2t-20m_c^2}{s} +\frac{3s+18r+20m_c^2}{t}+\frac{22rt-2t^2-38r^2-100rm_c^2+20tm_c^2}{s^2}\right. \nonumber\\
&&\left.-\frac{60rm_c^2+10sm_c^2}{t^2}-\frac{42r^2+40rm_c^2}{st} +\frac{18r^3+140r^2m_c^2}{s^2t}+\frac{140r^2m_c^2}{st^2}\right. \nonumber\\
&&\left. +\frac{3t^3-12r^3+18r^2t-12rt^2+40rtm_c^2-10t^2m_c^2-60r^2m_c^2}{s^3}+\frac{3r^4+40r^3m_c^2}{s^3t}\right.\nonumber\\
&&\left.-\frac{60r^3m_c^2}{s^2t^2}-\frac{10r^4m_c^2}{s^3t^2} \right)\nonumber\\
&&+\langle\frac{\alpha_sGG}{\pi}\rangle\frac{\sqrt{\lambda(s,t,r)\,\lambda(r,m_c^2,m_c^2)}}{1920\pi^4}\left(1-\frac{4m_c^2}{r}\right) \frac{1}{\sqrt{t\left(t-4m_c^2\right)}}\nonumber\\
&&\left(s+6r-14t+28m_c^2+\frac{26t^2-14r^2-116rt-52tm_c^2+232rm_c^2-40m_c^4}{s} \right. \nonumber\\
&&\left.+\frac{40m_c^4-12rm_c^2-2sm_c^2}{t}+\frac{28r^2m_c^2-80rm_c^4}{st}-\frac{120rm_c^4+20sm_c^4}{t^2} +\frac{280r^2m_c^4}{st^2}\right.\nonumber\\
&&\left.+\frac{6r^3-14t^3-26r^2t+34rt^2+28t^2m_c^2+52r^2m_c^2-68rtm_c^2+40tm_c^4-200rm_c^4}{s^2}\right.\nonumber\\
&&\left.+\frac{(t-r)^4-2t^3m_c^2+8r^3m_c^2-12r^2tm_c^2+8rt^2m_c^2+80rtm_c^4-20t^2m_c^4-120r^2m_c^4}{s^3}\right.\nonumber\\
&&\left.+\frac{280r^2m_c^4-12r^3m_c^2}{s^2t}+\frac{80r^3m_c^4-2r^4m_c^2}{s^3t}-\frac{120r^3m_c^4}{s^2t^2}-\frac{20r^4m_c^4}{s^3t^2}\right)\, ,
 \end{eqnarray}
where $\lambda(a,b,c)=a^2+b^2+b^2-2ab-2bc-2ca$, the $T^2$ is the Borel parameter. In calculations, we observe that there appears divergence due to the endpoint $t=4m_c^2$, we can avoid the endpoint divergence with the simple replacements $\frac{1}{t-4m_c^2} \to \frac{1}{t-4m_c^2+4m_s^2}$ and   $\frac{1}{\sqrt{t-4m_c^2}} \to \frac{1}{\sqrt{t-4m_c^2+4m_s^2}}$  by adding a small squared $s$-quark mass $4m_s^2=0.04\,\rm{GeV}^2$ \cite{Wang-Y4660-Decay}.
In this article, we take into account the perturbative terms and gluon condensate, which are vacuum expectations of the quark-gluon operators of the orders $\mathcal{O}(\alpha_s^0)$ and $\mathcal{O}(\alpha_s^1)$, respectively. In Refs.\cite{WangZG-IJMPA-Z3900,WangZG-Landau,WangZG-Hiddencharm}, we perform detailed  analysis, we observe that the two-meson scattering states cannot saturate the QCD sum rules for the tetraquark states and tetraquark molecular states,   the  tetraquark (molecular) states begin to receive contributions at the order $\mathcal{O}(\alpha_s^0/\alpha_s^1)$ rather than at the order $\mathcal{O}(\alpha_s^2)$.

We derive   Eq.\eqref{QCDSR} with respect to  $\tau=\frac{1}{T^2}$, then eliminate the
 pole residues $\lambda_{X/Y}$, and  obtain the QCD sum rules for
 the masses of the scalar, axialvector  and tensor fully-charm tetraquark states,
 \begin{eqnarray}\label{QCDSR-mass}
 M^2_{X/Y}&=&- \frac{\frac{d}{d \tau}\int_{16m_c^2}^{s_0} ds \int_{4m_c^2}^{\left(\sqrt{s}-2m_c\right)^2}dt \int_{4m_c^2}^{\left(\sqrt{s}-\sqrt{t}\right)^2}dr \, \rho(s,t,r)  \exp\left(-\tau s \right)}{\int_{16m_c^2}^{s_0} ds \int_{4m_c^2}^{\left(\sqrt{s}-2m_c\right)^2}dt \int_{4m_c^2}^{\left(\sqrt{s}-\sqrt{t}\right)^2}dr \, \rho(s,t,r)  \exp\left(-\tau s \right)}\, .
\end{eqnarray}

\section{Numerical results and discussions}

We take the standard value of the gluon condensate $\langle \frac{\alpha_sGG}{\pi}\rangle=0.012\pm0.004\,\rm{GeV}^4$
\cite{SVZ79,Reinders85,ColangeloReview}, and  take the $\overline{MS}$ mass $m_{c}(m_c)=(1.275\pm0.025)\,\rm{GeV}$
 from the Particle Data Group \cite{PDG}.
We take into account
the energy-scale dependence of  the  $\overline{MS}$ mass from the renormalization group equation,
 \begin{eqnarray}
m_c(\mu)&=&m_c(m_c)\left[\frac{\alpha_{s}(\mu)}{\alpha_{s}(m_c)}\right]^{\frac{12}{25}} \, ,\nonumber\\
\alpha_s(\mu)&=&\frac{1}{b_0t}\left[1-\frac{b_1}{b_0^2}\frac{\log t}{t} +\frac{b_1^2(\log^2{t}-\log{t}-1)+b_0b_2}{b_0^4t^2}\right]\, ,
\end{eqnarray}
  where $t=\log \frac{\mu^2}{\Lambda^2}$, $b_0=\frac{33-2n_f}{12\pi}$, $b_1=\frac{153-19n_f}{24\pi^2}$, $b_2=\frac{2857-\frac{5033}{9}n_f+\frac{325}{27}n_f^2}{128\pi^3}$,  $\Lambda=213\,\rm{MeV}$, $296\,\rm{MeV}$  and  $339\,\rm{MeV}$ for the flavors  $n_f=5$, $4$ and $3$, respectively  \cite{PDG}. In this article, we take the typical energy scale $\mu=2\,\rm{GeV}$ and choose the flavor number $n_f=4$ as we study the fully-charm tetraquark states.

We should choose suitable continuum threshold parameters $s_0$ to avoid contaminations from the first radial excited states and can borrow some ideas from the conventional charmonium states and the charmonium-like states.
The  masses of the ground state and the first radial excited state of the vector charmonium states are $M_{J/\psi}=3.0969\,\rm{GeV}$ and   $M_{\psi^\prime}=3.686097\,\rm{GeV}$ respectively from the Particle Data Group \cite{PDG}, the energy gap is $M_{\psi^\prime}-M_{J/\psi}=589\,\rm{MeV}$.
We usually assign the $Z_c(4430)$  to be the first radial excitation of the $Z_c(3900)$ according to the
analogous decays $Z_c^\pm(3900)\to J/\psi\pi^\pm$ and $Z_c^\pm(4430) \to \psi^\prime\pi^\pm$,
and the analogous mass gap  $M_{Z_c(4430)}-M_{Z_c(3900)}=591\,\rm{MeV}$ from the Particle Data Group \cite{PDG,Maiani-Z4430-1405}.
On the other hand, we can tentatively assign the $X(3915)$ and $X(4500)$  to be   the ground state and the first radial excited state of the axialvector-diquark-axialvector-antidiquark type scalar $cs\bar{c}\bar{s}$ tetraquark states  according to the energy gap $M_{X(4500)}-M_{X(3915)}=588\,\rm{MeV}$ \cite{PDG,Lebed-X3915,WangZG-X4500}.
  If the resonance  structure $Z_c(4600)$ have the $J^{PC}=1^{+-}$, we can tentatively assign  the $Z_c(4020)$ and $Z_c(4600)$ to be the ground state  and the first radial excited state of the axialvector-diquark-axialvector-antidiquark type or scalar-diquark-axialvector-antidiquark type axialvector tetraquark states respectively   considering the   energy gap  $M_{Z_c(4600)}-M_{Z_c(4020)}=576\,\rm{MeV}$  \cite{PDG,WangZG-Hiddencharm,ChenHX-Z4600-A}.

Now we can obtain the conclusion tentatively that the energy gaps between the ground states and the first radial excited states
of the hidden-charm tetraquark states are about $585\,\rm{MeV}$. In the present work,  we can  choose the continuum threshold parameters   $\sqrt{s_0}= M_{S/Y/T}+0.55\,\rm{GeV}$ as a constraint tentatively and vary the continuum threshold parameters and Borel parameters to satisfy
the  two basic    criteria of the QCD sum rules, the ground state  dominance  at the hadron  side and the operator product expansion  converges at the QCD side.

 After  trial and error,    we  obtain the reasonable  continuum threshold parameters and  Borel parameters, which are shown in Table \ref{Borel}. In the Borel windows, the pole contributions or ground state contributions are about $(39-62)\%$, the central values are larger than $50\%$,  the pole dominance at the hadron side is well satisfied. On the other hand,  the dominant contributions come from the perturbative terms in the Borel windows,  the operator product expansion  converges  very well.

\begin{figure}
 \centering
 \includegraphics[totalheight=5cm,width=7cm]{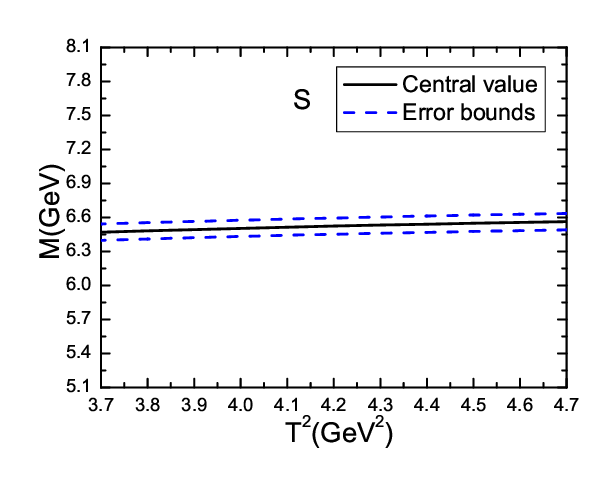}
 \includegraphics[totalheight=5cm,width=7cm]{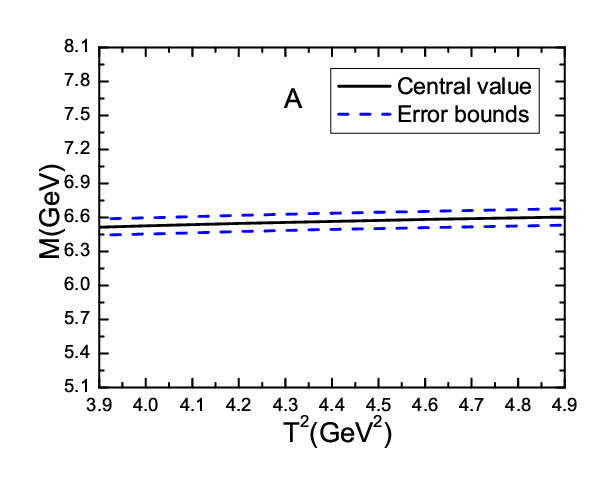}
 \includegraphics[totalheight=5cm,width=7cm]{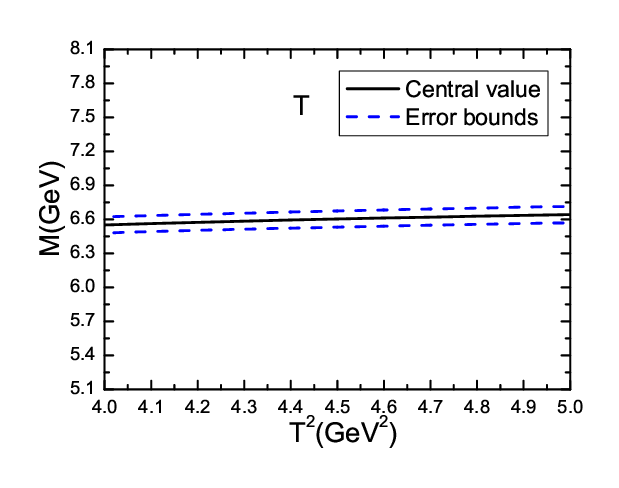}
         \caption{ The masses   of the fully-charm  tetraquark states  with variations of the Borel parameters $T^2$,  where the $S$, $A$ and $T$ denote the scalar, axialvector and tensor tetraquark states, respectively.  }\label{mass-cccc}
\end{figure}

Now let us  take into account all uncertainties of the input parameters, such as the continuum threshold parameter, the $c$-quark mass, the gluon condensate, the Borel parameter,  and obtain the values of the masses and pole residues of the scalar, axialvector and tensor fully-charm tetraquark states, which are  shown explicitly in Table \ref{Borel} and Fig.\ref{mass-cccc}. From Fig.\ref{mass-cccc}, we can see that the predicted masses  are rather stable with variations of the Borel parameters, the uncertainties originate from the Borel parameters  are very small, it is reliable to extract the tetraquark masses.

The quantum field theory allows  non-vanishing couplings between an interpolating current and a hadron (or several hadrons or a hadron system or several hadron systems) provided they have the same quantum numbers. On the other hand, a hadron  has  many Fock states, a Fock state couples potentially to an interpolating current with the same quantum numbers.
In the present case, we can introduce the mixing angle $\theta$ and study the fully-charm tetraquark states with the currents $AA\cos\theta+\tilde{A}\tilde{A}\sin\theta$, where the $AA$ and $\tilde{A}\tilde{A}$ denote the $AA$-type and $\tilde{A}\tilde{A}$-type tetraquark currents respectively. We expect to obtain better QCD sum rules via fine-turning the parameter $\theta$ as the fully-charm tetraquark states have more than one Fock states.

\begin{table}
\begin{center}
\begin{tabular}{|c|c|c|c|c|c|c|c|}\hline\hline
$J^{PC}$       &$T^2(\rm{GeV}^2)$    &$\sqrt{s_0}(\rm{GeV})$  &pole          &$M_{X/Y}(\rm{GeV})$   &$\lambda_{X/Y}(10^{-1}\rm{GeV}^5)$ \\ \hline

$0^{++}$       &$3.9-4.5$            &$7.05\pm0.10$           &$(39-63)\%$   &$6.52\pm0.10$         &$6.17\pm1.34$  \\ \hline

$1^{+-}$       &$4.1-4.7$            &$7.10\pm0.10$           &$(38-62)\%$   &$6.57\pm0.10$         &$5.17\pm1.08$  \\ \hline

$2^{++}$       &$4.2-4.8$            &$7.15\pm0.10$           &$(39-62)\%$   &$6.60\pm0.10$         &$7.95\pm1.63$  \\ \hline\hline

\end{tabular}
\end{center}
\caption{ The Borel parameters, continuum threshold parameters,   pole contributions, masses and pole residues of the fully-charm   tetraquark states. }\label{Borel}
\end{table}

\begin{table}
\begin{center}
\begin{tabular}{|c|c|c|c|c|c|c|c|}\hline\hline
$J^{PC}$                              &1S                    &2S                 &3S                &$\overline{1\rm S}$   &$\overline{2\rm S}$     \\ \hline

$0^{++}(\tilde{A}\tilde{A})$          &$6.52\pm0.10$         &                   &                  &$0.33\pm0.10$         &                         \\

$0^{++}(AA)$                          &$5.99\pm0.08$         &$6.48\pm0.08$      &$6.94\pm0.08$     &$-0.20\pm0.08$        &$0.29\pm0.08$           \\ \hline

$1^{+-}(\tilde{A}\tilde{A})$          &$6.57\pm0.10$         &                   &                   &$0.38\pm0.10$         &                         \\

$1^{+-}(AA)$                          &$6.05\pm0.08$         &$6.52\pm0.08$      &$6.96\pm0.08$      &$-0.14\pm0.08$        &$0.33\pm0.08$          \\ \hline

$2^{++}(\tilde{A}\tilde{A})$          &$6.60\pm0.10$         &                   &                   &$0.41\pm0.10$         &                          \\

$2^{++}(AA)$                          &$6.09\pm0.08$         &$6.56\pm0.08$      &$7.00\pm0.08$      &$-0.10\pm0.08$        &$0.37\pm0.08$         \\ \hline

\hline

\end{tabular}
\end{center}
\caption{ The predicted fully-charm  tetraquark masses from the QCD sum rules, where the $AA$-type tetraquark masses are taken from Refs.\cite{WZG-QQQQ-EPJC,WZG-QQQQ-CPC}, the overline on the 1S and 2S denotes that the $J/\psi J/\psi$ threshold is subtracted. }\label{cccc-mass-spectrum}
\end{table}

 In Table \ref{cccc-mass-spectrum}, we also present the masses of the ground states and the first radial excited states of the $AA$ type tetraquark states from the QCD sum rules \cite{WZG-QQQQ-EPJC,WZG-QQQQ-CPC}, and the masses of the  second radial excited states of the $AA$ type tetraquark states from the  Regge trajectories \cite{WZG-QQQQ-CPC}. From the Table, we can see that the 1S $\tilde{A}\tilde{A}$ type tetraquark states and the 2S $AA$-type tetraquark states have almost degenerated masses, they lie about $0.35\pm 0.09\,\rm{GeV}$ above the $J/\psi J/\psi$ threshold, the broad structure above the $J/\psi J/\psi$ threshold observed by the LHCb  collaboration maybe consist of several
 diquark-antidiquark type $cc\bar{c}\bar{c}$ tetraquark states, more precise measurements are still needed,  while the narrow structure $X(6900)$ can be assigned to be  the  second radial excited state of the scalar or axialvector $cc\bar{c}\bar{c}$ tetraquark state \cite{WZG-QQQQ-CPC}.

  In the present work, we introduce the explicit P-wave to construct the diquark operators therefore the tetraquark operators.  Without introducing the explicit P-waves, we can take the diquark operators $Q^T_j C Q_k$,  $Q^T_j C\gamma_5 Q_k$ and  $Q^T_j C\gamma_\mu\gamma_5 Q_k$ in the symmetric color sextet and the diquark operators $Q^T_j C\gamma_\mu Q_k$, $Q^T_j C\sigma_{\mu\nu} Q_k$,  $Q^T_j C\sigma_{\mu\nu}\gamma_5 Q_k$ in the antisymmetric color antitriplet, which satisfy the Fermi-Dirac statistics,  as the basic constituents  to construct the fully-heavy tetraquark states with the same flavor.   In Ref.\cite{WZG-QQQQ-EPJC}, we choose the diquark operators $\varepsilon^{ijk}Q^T_j C\gamma_\mu Q_k$ as the basic constituents to study the ground state masses of the $J^{PC}=0^{++}$, $1^{+-}$ and $2^{++}$ fully heavy tetraquark states, and obtain the lowest mass $6.0\,\rm{GeV}$ for the $cc\bar{c}\bar{c}$ tetraquark states. In Ref.\cite{WZG-QQQQ-CPC}, we study the masses of the first radial excited states of the $cc\bar{c}\bar{c}$ tetraquark states with the QCD sum rules and obtain masses of the second radial excited states with the Regge trajectories, the predicted masses of the first radial excited states and the second radial excited states are about $6.5\,\rm{GeV}$ and $6.9\,\rm{GeV}$, respectively, see Table \ref{cccc-mass-spectrum}. In Ref.\cite{ChenW-PLB-2017}, W. Chen et al take both the diquark operators in the color sextet and color antitriplet as the basic constituents to study the mass spectrum of the fully heavy tetraquark states with the moments QCD sum rules, and obtain the tetraquark masses about $6.3\sim 6.9\,\rm{GeV}$, where the lowest mass  of the scalar $cc\bar{c}\bar{c}$ tetraquark states is about $6.4\,\rm{GeV}$. In Ref.\cite{ZhangJR-QQQQ}, J. R. Zhang takes $c^T_j C c_k$,  $c^T_j C\gamma_5 c_k$ and  $c^T_j C\gamma_\mu\gamma_5 c_k$ in the  color sextet and the diquark operator $c^T_j C\gamma_\mu c_k$ in the  color antitriplet to study  the $J^P=0^+$ fully-charm tetraquark states with the QCD sum rules, and obtain almost degenerated masses, about $6.5\,\rm{GeV}$. Those  different predictions in Refs.\cite{WZG-QQQQ-EPJC,ChenW-PLB-2017,WZG-QQQQ-CPC,ZhangJR-QQQQ} maybe originate from the different input parameters and different Borel windows. They are all compatible with the experimental data from the LHCb collaboration within uncertainties at the present time \cite{LHCb-cccc-2006}, more experimental data are still needed to select the best QCD sum rules.

\section{Conclusion}
In this article, we introduce a relative P-wave to construct the doubly-charm axialvector diquark operator, then take the doubly-charm axialvector (anti)diquark operator as the basic constituent to construct the scalar and tensor tetraquark currents to study the scalar, axialvector and tensor fully-charm tetraquark states with the QCD sum rules.  The numerical results indicate that  the ground state  $\tilde{A}\tilde{A}$ type tetraquark states and the first radial excited states of the $AA$ type tetraquark states  have almost  degenerated masses, they lie about $0.35\pm 0.09\,\rm{GeV}$ above the $J/\psi J/\psi$ threshold, the broad structure above the $J/\psi J/\psi$ threshold observed by the LHCb  collaboration maybe consist of several
 diquark-antidiquark type fully-charm  tetraquark states.

\section*{Acknowledgements}
This  work is supported by National Natural Science Foundation, Grant Number  11775079.

\end{document}